\documentclass{jfm}
\usepackage{epstopdf, epsfig}
\usepackage{siunitx}
\usepackage[utf8]{inputenc}
\usepackage{tikz}

\usepackage{pgfplots}
\pgfplotsset{compat=1.14}

\newlength\figH
\newlength\figW
\begin{document}

\newtheorem{lemma}{Lemma}
\newtheorem{corollary}{Corollary}

\shorttitle{Fingering instability of a  stretched liquid bridge}
\shortauthor{S. Brulin, I. V. Roisman, and  C. Tropea}

\title{Fingering instability of a viscous liquid bridge stretched by an accelerating substrate}

\author{Sebastian Brulin, Ilia V. Roisman \corresp{\email{roisman@sla.tu-darmstadt.de}} \and  Cameron Tropea  }

\affiliation{Technische Universit\"at Darmstadt, Germany, Institute for Fluid Mechanics and Aerodynamics, Alarich-Weiss-Stra\ss e 10, 64287 Darmstadt, Germany}

\maketitle

\begin{abstract}
When a liquid viscous bridge between two parallel substrates is stretched by accelerating one substrate, its interface recedes in the radial direction. In some cases the interface becomes unstable. Such instability leads to the emergence of a network of fingers. In this study the mechanisms of the fingering are studied experimentally and analysed theoretically. 
The experimental setup allows a constant acceleration of the movable substrate with up to 180 m/s$^2$. The phenomena are observed using two high-speed video systems. The number of fingers is measured for different liquid viscosities and liquid bridge sizes. 
A  linear stability analysis of the bridge interface takes into account the inertial, viscous and capillary effects in the liquid flow. The theoretically predicted maximum number of fingers, corresponding to a mode with the maximum amplitude, and a threshold for the onset of finger formation are proposed. Both models agree well with the experimental data up to the onset of emerging cavitation bubbles. 

\end{abstract}


\section{Introduction}

The phenomena of fluid bridge stretching have first been studied by \citet{Plateau1864}, \citet{stefan_versuche_1875} and \citet{Rayleigh} more than one hundred years ago. Since then the dynamics of liquid jets and bridges have been studied extensively. Several comprehensive reviews of this field present state-of-the-art modeling approaches \citep{Yarin1993,VillermauxARFM,Villermaux,eggers1997nonlinear,schulkes1993dynamics}. 

Liquid jet or liquid bridge stretching is a phenomenon relevant to many practical applications like rheological measurements, atomisation, crystallization, car soiling, oil recovery and typical industrial printing processes such as gravure, flexography, lithography or roll coating \citep{jarrahbashi2016early,ambravaneswaran1999effects,marmottant2004spray,gordillo2010generation,gaylard_surface_2017,xu_study_2017}. The dynamics of liquid bridges also govern coalescence processes of solid wetted particles \citep{cruger2016coefficient}.

A large class of models have been developed for relatively long jets. The long-wave approach for a nearly cylindrical jet describes well, for example, the transverse instability the jet exposed to an air flow, introduced by \citet{entov1984dynamics}. A slender jet model was also used by \citet{eggers_drop_1994} and \citet{Eggers} to show that liquid bridge pinching is universal, but asymmetric in the pinching region for pinned liquid bridges. \citet{Papageorgiou} later introduced an alternative model with a symmetrical geometry solution in the pinching region. A more recent study  from \citet{qian_motion_2011} has shown the effects of surface wettability and moving contact lines on liquid bridge break-up behaviour for stretching speeds up to $\SI{600}{\mu m/s}$. Further studies use this long-wave approach to show the importance of the pinching position for the break-up time \citep{yildirim_deformation_2001} or the behaviour of non-Newtonian liquids on bridge thinning and break-up behaviour \citep{anna_elasto-capillary_2000, mckinley_visco-elasto-capillary_nodate}. 

If the height of a liquid bridge ($H_0$) is much smaller than its diameter ($D_0$) the dimensionless height is $\lambda \ll 1$ with $\lambda =H_0/D_0$, therefore the modeling approach has to be different from previous studies. For such cases the surface of the liquid bridge can become unstable because of the high interface retraction rates and small initial liquid bridge heights. Due to the small initial heights and wide initial diameter, the conditions are similar those in Hele-Shaw flow cells.

A frequently observed phenomenon are finger patterns formed from growing instabilities in fixed height Hele-Shaw cells for transverse \citep{saffman_penetration_1958} or radial flows \citep{mccloud_experimental_1995,mora_saffman-taylor_2009}. The study from \cite{maxworthy_experimental_1989} compares modified wavenumber theories based on fastest growing modes from \cite{park_gorell_homsy_1984} and \cite{Schwartz} to experiments in radial flows. Since the measurements of this study are performed with constant acceleration and a maximum amplitude criteria is used, which is explained later, the wavenumbers are not comparable. \cite{paterson_radial_1981} derived a prediction for the number of  fingers formed at a radially expanding interface of the liquid spreading between two fixed substrates.

The problem is completely different if the flow is caused by the motion of the substrates and the gap thickness changes in time. One of the examples is the study by \cite{ward2011capillary} in which the rate equation for the displacement of a liquid bridge under a defined pulling force is investigated. 

The measurements from \citet{amar_fingering_2005} were conducted at very low stretching speeds of $20-\SI{50}{\mu m/s}$, high viscosities \SI{30}{Pas} and large initial heights. For a lifting Hele-Shaw cell \citet{nase_dynamic_2011} developed a model for the interfacial stability of the liquid, leading to the prediction of the maximum number of the fingers.
\cite{dias_determining_2013,shelley_hele_1997,dutta_radially_2003,amar_fingering_2005, spiegelberg_stress_1996} studied the liquid bridge stretching in the lifted Hele-Shaw setups. In most of these cases the stretching speed is constant and is relatively small such that the inertial effects are comparably small.

The analysis of the finger instability has been further generalized by \citet{dias_determining_2013}, where the influence of the radial viscous stresses at the meniscus has been taken into account. It has been shown that for identification of the most unstable mode, the maximum amplitude has to be considered instead of the usual approach of selecting the fastest growing modes also by \citet{dias2013wavelength}. This approach accounts for the non-stationary effects in the flow even if the substrate velocity is constant. The amplitude growth due to the disturbances is not exponential since the parameters of the problem, mainly the thickness of the gap, change in time. More recently \citet{anjos_inertia-induced_2017} showed in a analytical and numerical study that inertia has a significant impact on the finger formation at higher velocities, especially on the dendritic like structures on the finger tips.

In this study the flow in a thin liquid bridge between two substrates, generated by an accelerating downward motion of the lower plate, is studied experimentally and modelled theoretically. This situation is a generic model for processes like gravure printing or water splash due to a tire rolling on a wet road. The main feature of this case is the variation of the velocity of the substrate lifting with relatively high acceleration. The acceleration of the substrate in the present study can reach 180 m/s$^2$. The stretching of the liquid bridge between these two substrates is observed using a high-speed video system. 
In the experimental part of the study the outcome of the liquid bridge stretching is characterized and the number of the fingers is measured. 

A linear stability analysis of the liquid bridge accounts for the inertial term, viscous stresses and capillary forces in the liquid flow. The analysis allows prediction of the maximum number of fingers. This number is obtained from the condition of the maximum amplitude of the corresponding mode due interface perturbations. The second criteria for the finger threshold is associated with the limiting value of the dimensionless wave amplitude. Both criteria lead to the same scaling of the threshold parameter for fingering and agree well with the experimental observations.

It should be noted that in this study the number of fingers is predicted as a function of the substrate acceleration. Therefore it is not possible to directly compare with the previous studies based on a constant substrate velocity. There is no characteristic velocity in the present study, only acceleration.

\section{Experimental Method}\label{sec:exp}
\subsection{Experimental setup and procedure}
\begin{figure}
    \centering
    a)\includegraphics[width=.5\textwidth]{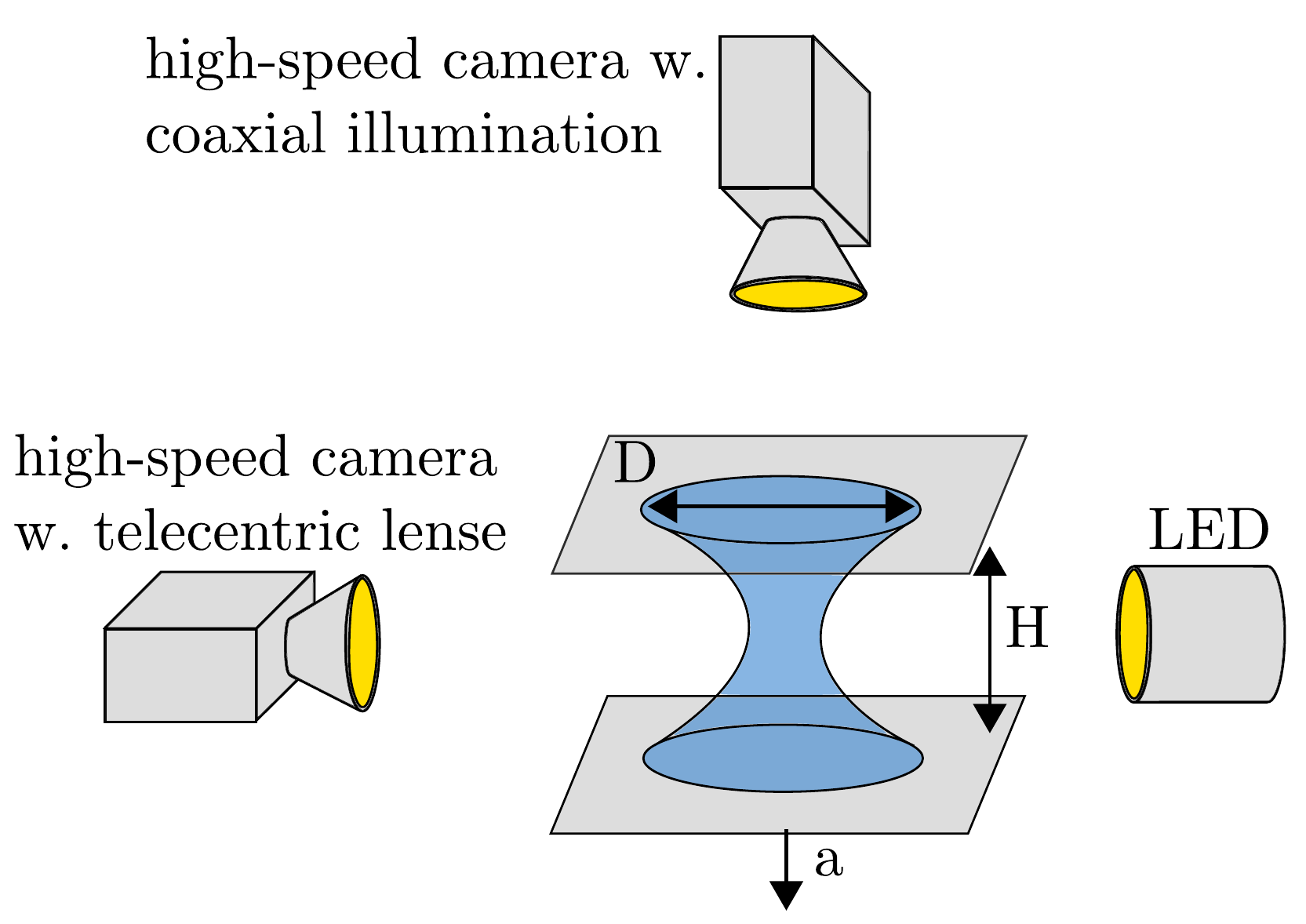}\hspace{1.5 cm}    b)\includegraphics[width=.22\textwidth]{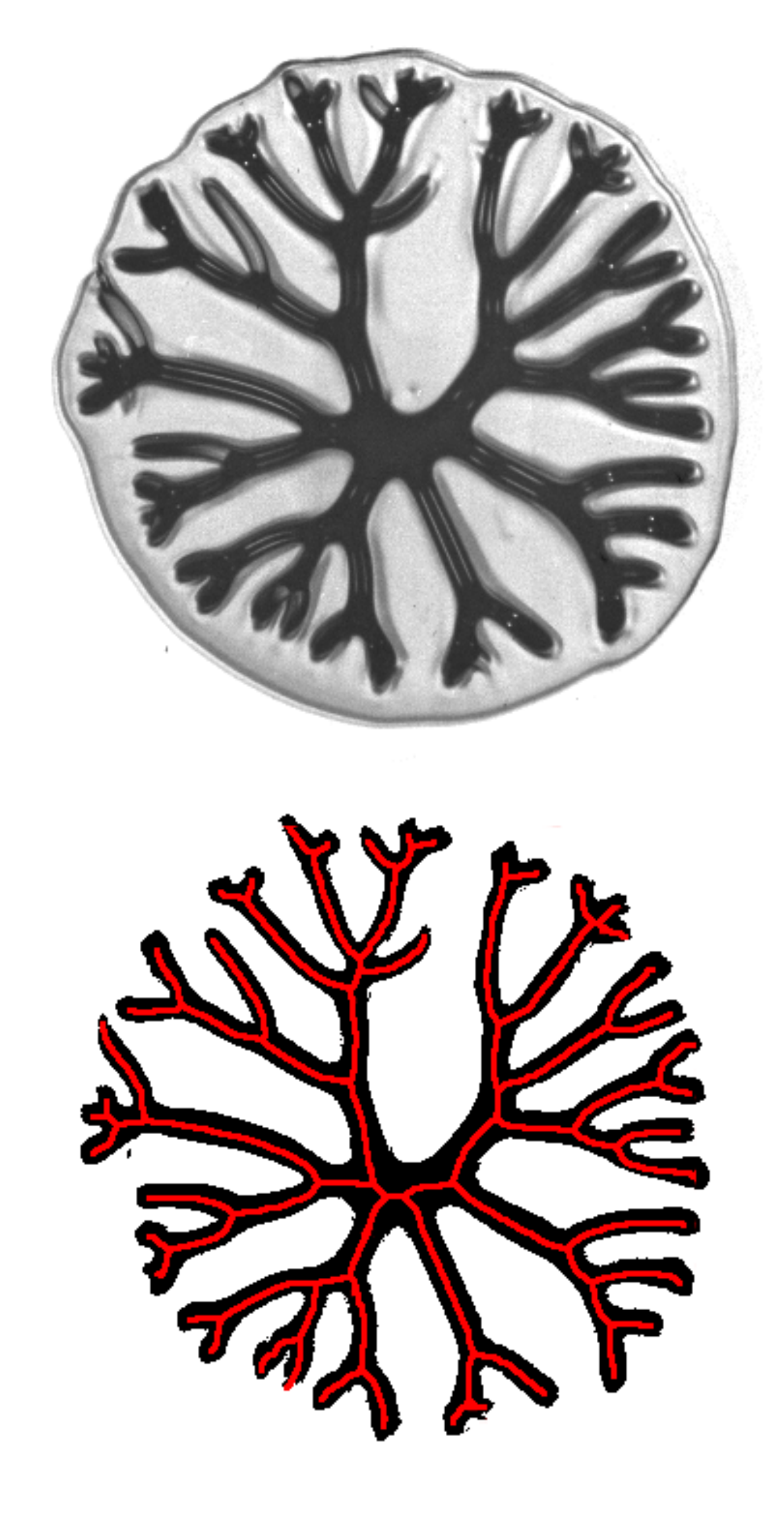}
    \caption{Experimental method: a) the setup and b) post-processing of images.}
    \label{fig:Experimental}
\end{figure}

The experimental setup for stretching a liquid bridge is shown schematically in fig.~\ref{fig:Experimental}a).
The stretching system consists of two substrates orientated horizontally. The lower substrate is mounted on a linear drive which allows accelerations from $\SI{10}{m/s^2}$ to $\SI{180}{m/s^2}$. The position accuracy of the linear drive is about $\SI{5}{\mu m}$. The upper substrate is fixed. 
Both substrates are  transparent, fabricated from glass with a roughness of $R_a = \SI{80}{nm}$. The initial thickness of the gap, $H_0$, in the experiments varies from 20 to 140 $\mu$m, as shown below in fig. \ref{fig:instability_sketch}.

 A microliter syringe is used as a fluid dispensing system. To investigate the effect of the liquid properties, two water-glycerol mixtures  with different viscosities are used: \textit{Gly50} with the viscosity $\mu=5.52\times 10^{-3}$ \si{\pascal\, \second}, surface tension $\sigma=67.3\times 10^{-3}$ \SI{}{N\ m^{-1}}  and density $\rho=1129.$ \si{\kilogram\ \metre^{-3}}; and \textit{Gly80} whose properties are  $\mu=5.36\times 10^{-2}$ \si{\pascal\, \second}, $\sigma=65.5\times 10^{-3}$ \SI{}{N\ m^{-1}}  and  $\rho=1211.$ \si{\kilogram\ \metre^{-3}}. The possible variation of the liquid properties with temperature has been accounted for in the data analysis. 
 
 To examine the size effects on the outcome of the liquid bridge stretching, the liquid volumes of the bridge are varied between $1$ and $\SI{5}{\mu l}$. This allows variation of the initial height-to-diameter ratio, $\lambda = H_0/D_0$, in the range $0.003<\lambda<0.2$. 
 
 The observation system consists of two high-speed video systems. The side view camera is equipped with a telecentric lens. The camera on top uses a 12x zoom lens system. The images are captured with a resolution of one megapixel at a frequency of \SI{12.5}{kfps}.

The post-processing for the top view was performed with the help of trainable weka \citep{arganda-carreras_trainable_2017}, a machine learning algorithm assisting with the segmentation of the images. Afterwards the images were skeletonized and the number of  fingers was counted. An example of the segmentation and skeleton is shown in fig.~\ref{fig:Experimental}b) and later in fig. \ref{fig:functionRR0} the number of fingers is plotted over $R/R_0$.

\subsection{Observations of the bridge stretching}
\begin{figure}
    \centering
    a)\includegraphics[width=.42\textwidth]{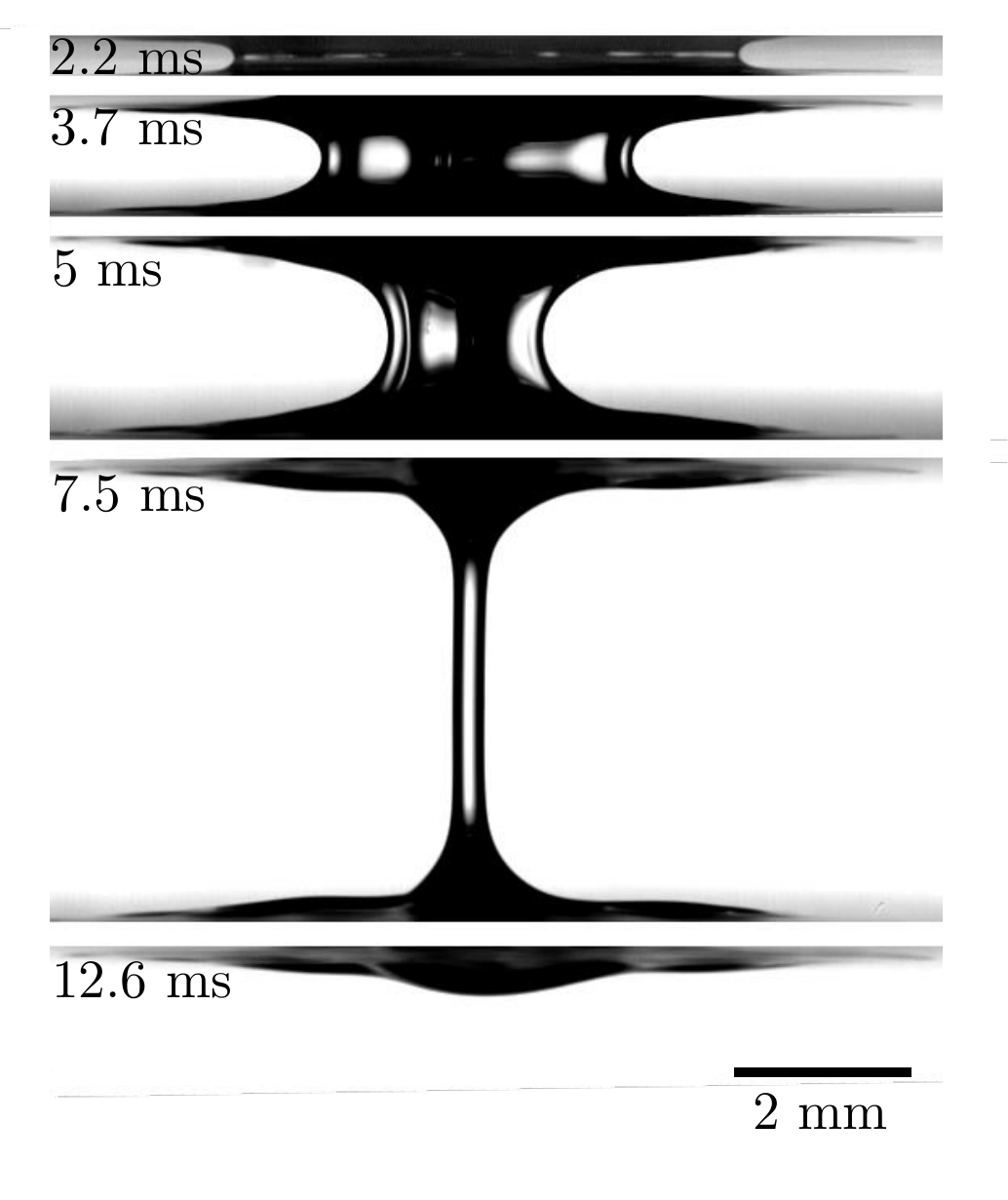}
    b)\setlength{\figW}{5.2 cm}
	\setlength{\figH}{5.2 cm}\input{plots/HH0_DmD0.tex} 
    \caption{Evolution of the diameter of a liquid bridge. (a) Side views of a  Gly80 bridge stretched with the constant acceleration \SI{180}{m/s^2}. The initial gap is $\SI{20}{\mu m}$ and the gap-to-diameter ratio is $\lambda = 0.02$. (b) The scaled bridge middle diameter $D_M/D_0$ as a function of the dimensionless gap width $H/H_0$ for various substrate accelerations. The curve corresponds to the predictions based on (\ref{eq:Roft}).} 
    \label{fig:instability_growth_side}
\end{figure}

\begin{figure}
   \centering
\begin{tabular}{cccc}
  \hspace{0.3cm}\includegraphics[height=.7\textheight]{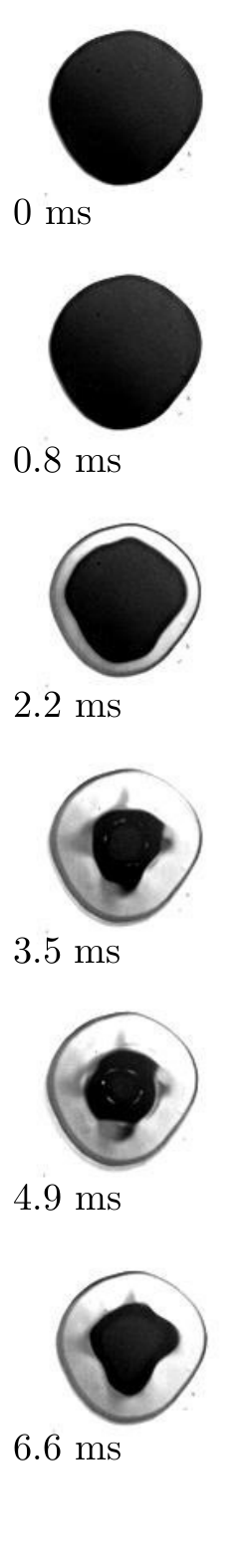}\hspace{0.3cm} & \hspace{0.3cm}\includegraphics[height=.7\textheight]{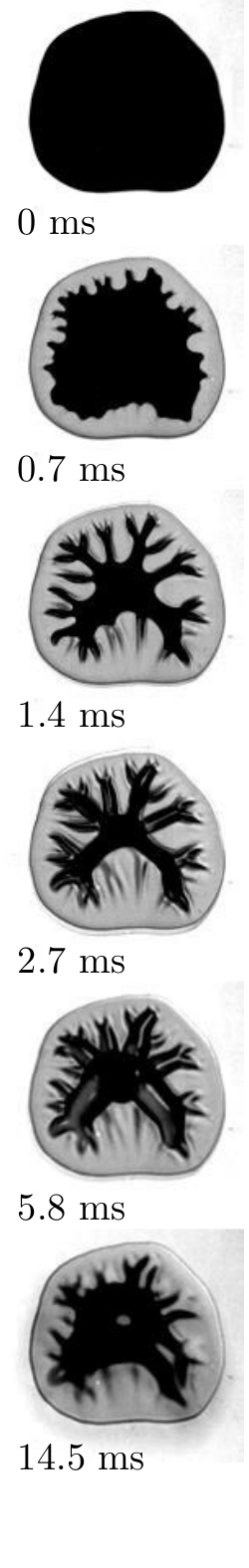} \hspace{0.3cm}&\hspace{0.3cm} \includegraphics[height=.7\textheight]{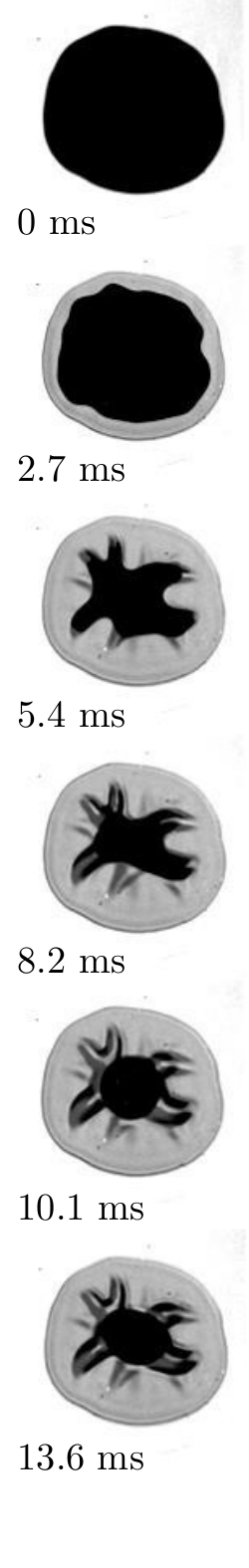}\hspace{0.3cm} &\hspace{0.3cm}\includegraphics[height=.7\textheight]{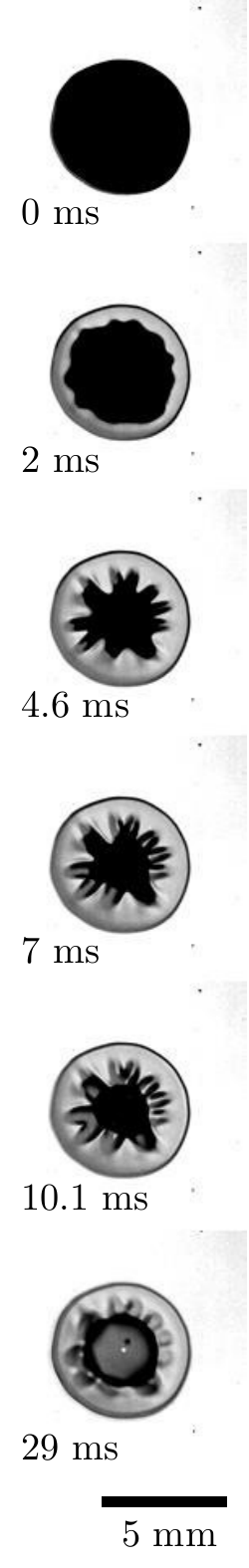}\hspace{0.3cm}\\
  a) & b) & c) & d) \\
\end{tabular}
    \caption{Top view on the receding interface due to the bridge stretching at various experimental conditions. a - liquid $Gly50$, substrate acceleration $a=$ \SI{180}{m/s^2}, relative gap width $\lambda=0.03$; b - $Gly50$, $a=$ \SI{180}{m/s^2}, $\lambda=0.006$; c - $Gly50$, $a=$ \SI{10}{m/s^2}, $\lambda=0.06$; d - $Gly80$, $a=$ \SI{10}{m/s^2}, $\lambda=0.03$.}    \label{fig:bottom_view}
\end{figure}

The experimental setup allows shadowgraphy images of the contact area between the substrate and liquid to be captured during the stretching process. An example of the side-view, high-speed visualization of a stretching $Gly80$ bridge is shown in  fig.~\ref{fig:instability_growth_side}a). In this example the substrate acceleration is $\SI{180}{m/s^2}$ and the initial height is $\SI{20}{\mu m}$. The initial liquid bridge height-to-diameter ratio is $\lambda=0.02$. 
While the middle diameter, $D_M$, of the bridge reduces during the stretching process, a thin liquid film remains on both substrates. The contact line stays pinned for all performed experiments, evident from the top views from fig. \ref{fig:bottom_view}. After $\SI{12.6}{ms}$ the bridge  pinches off. In fig.~\ref{fig:instability_growth_side}b) the evolution of the scaled bridge diameter during stretching is shown as a function of the dimensionless gap width. For the stage when $H \ll D_M $ the evolution of the bridge diameter is universal. It does not depend on the substrate acceleration or liquid properties. 

Several typical top views of the liquid bridge through the transparent substrate are shown in fig.~\ref{fig:bottom_view} at different instants for various experimental parameters. In some cases the onset of instability can be clearly seen, which leads to the appearance of a net of fingers. The most stable case in fig.~\ref{fig:bottom_view}a corresponds to the relatively wide gap and low acceleration. The most unstable case, associated with the highest number of fingers, corresponds to the highest accelerations and smaller initial gap widths, as in the example fig.~\ref{fig:bottom_view}b. In the example fig.~\ref{fig:bottom_view}c fingers can be observed even with relatively small substrate acceleration, but for small dimensionless heights $\lambda$. Fig.~\ref{fig:bottom_view}d shows how increased liquid viscosity compared to fig.~\ref{fig:bottom_view} leads to an evolved finger pattern.

Increasing the  substrate acceleration or viscosity enhance the fingering instability, whereas with an increasing dimensionless height $\lambda$ the finger formation is mitigated. 

\section{Stability analysis of the bridge interface}
In this study a stability analysis is performed on the basis of experimental measurements of the flow in a thin gap between two substrates. The problem is linearized in the framework of the long-wave approximation. 

\subsection{Basic flow}

The flow field in the stretching liquid bridge can be subdivided into two main regions: meniscus region and the central, inner region, which is not influenced by the meniscus. The solution for an axisymmetric creeping flow between two parallel substrates, one of which moves, is well known \citep{Landau1959}. The axial and the radial components of the velocity field are
\begin{equation}\label{velocBaseInner}
u_{0,r} = -\frac{3 \dot H r z}{H^2} \left(1-\frac{z}{H}\right), \quad  u_{0,z} = \frac{3\dot H z^2}{H^2} \left(1-\frac{2 z}{3 H}\right).
\end{equation}
This velocity field satisfies the equation of continuity, the momentum balance equation and the kinematic conditions at both substrates. Unfortunately this solution is not applicable to the case when the effect of the substrate acceleration becomes significant. 
Moreover, the expression for the velocity field between two substrates (\ref{velocBaseInner}) is not applicable at the interface of the meniscus. It does not satisfy the conditions for the pressure at the interface,  determined by the Young-Laplace equation; and it does not satisfy the conditions of zero shear stress at this interface. Moreover, this velocity field is not able to accurately predict the rate of change of the meniscus radius $\dot R$. Let us assume the rate change of the minimum meniscus radius at the middle plane as $\dot R = u_{0,r}$ at $z= H/2$. Solution of the equation for the meniscus propagation with the help of (\ref{velocBaseInner}) yields $R=R_0 (H_0/H)^{3/16}$. This solution does not agree with the experimental data for the evolution of the meniscus radius, therefore the flow in the meniscus region has to be treated differently. 

\begin{figure}
    \centering
    \includegraphics[width=.8\textwidth]{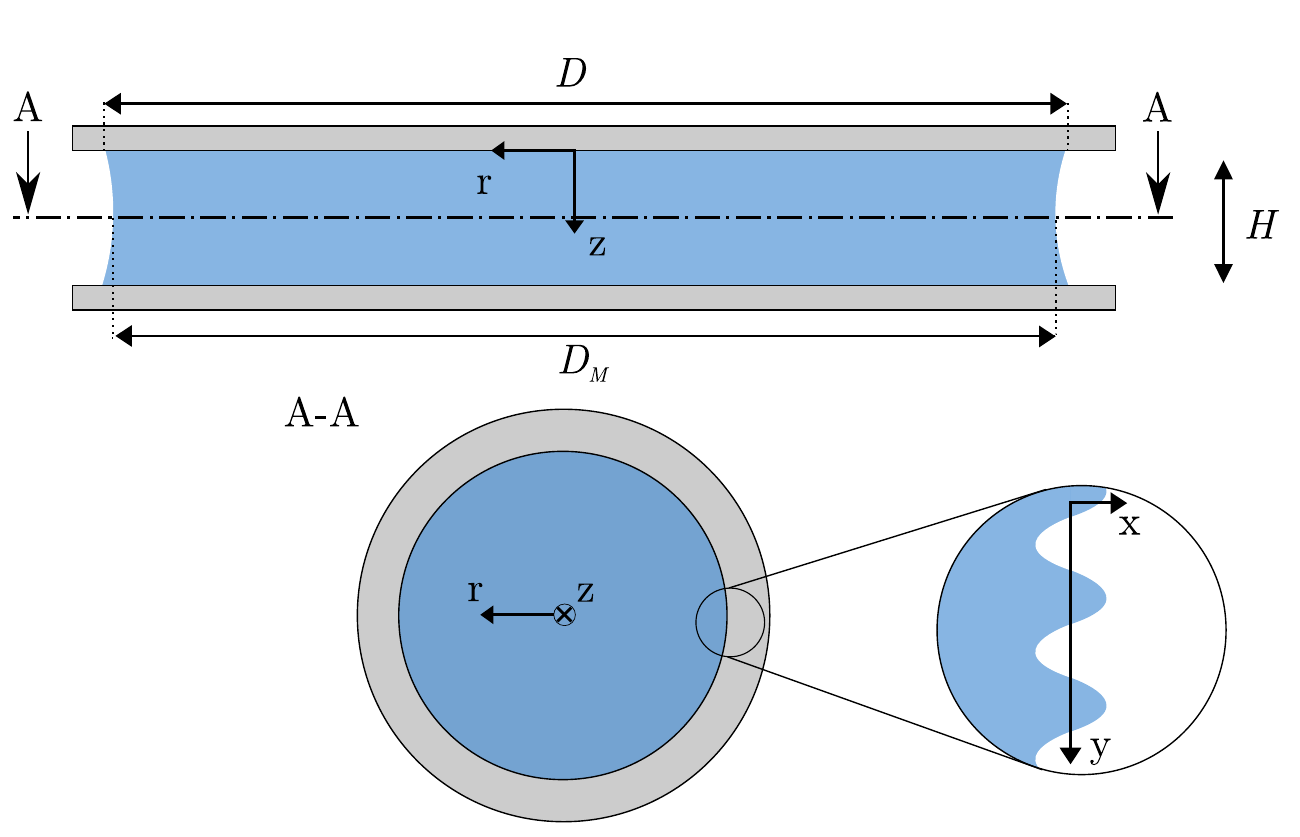}    
    \caption{A cylindrical frame of reference is used at the symmetry axis ($r, z$). The area of interest for instability analysis is magnified and a Cartesian frame of reference is used at the interface ($\{x, y\}$).}
    \label{fig:instability_sketch}
\end{figure}
An expression for the radius and the height can be derived from the overall mass balance. Where the initial thickness is $H_0$, the initial radius is $R_0$ and the lower substrate moves with a constant acceleration $a$. The radius of the bridge $R(t)$ of the bridge meniscus can be estimated as
\begin{equation}\label{eq:Roft}
    R(t)=R_0 \left[\frac{H_0}{H(t)}\right]^{1/2}, \quad H(t) = H_0 + \frac{a t^2}{2}
\end{equation}
which gives an valid estimate for the initial times when $R(t)\gg H(t)$, as demonstrated on the graph in fig.~\ref{fig:instability_growth_side}b).

The flow in the meniscus region has to be treated separately. This flow has to satisfy the boundary conditions at the curved meniscus interface and must include also the corner flows \citep{moffatt1964viscous,anderson1993two}. The model of the meniscus flow is not trivial and can lead to multiple solutions \citep{gaskell1995modelling}. However an accurate solution for the meniscus stability problem has to be based on the  meniscus velocity field, since the stresses in this region actually govern the meniscus instability. 

In this study we assume that the main reason for the meniscus instability is the appearance of a normal pressure gradient at the meniscus interface. This mechanism is analogous to the Rayleigh-Taylor instability, where the pressure gradient is caused by gravity or by the interface acceleration. This approximate solution is valid only for the case of very small relative gap thickness, $\lambda \ll 1$. Note also, that the ratio of the axial and radial components of the liquid velocity is comparable with $\lambda$. The stresses  associated with the axial flow are thus much smaller than those associated with the radial velocity component. 

Consider only the dominant terms of the pressure gradient at the interface. The pressure gradient includes the viscous stresses and the inertial terms associated with the material acceleration of the meniscus $\ddot R$. The approximation is based on the fact that the radial velocity in the liquid at the interface at the middle plane $(z=H/2)$ is equal to $\dot R$. The value of the pressure gradient is then estimated from the Navier-Stokes equations with the help of (\ref{eq:Roft}) in the form
\begin{equation} \label{eq:dpdrAcc}
      p_{0,r}= - b \mu \frac{\dot R}{H(t)^2} - \rho \ddot R
      = \mu a t b\frac{ \sqrt{H_0} R_0}{2 H^{7/2}}+\rho a \frac{ \sqrt{H_0} R_0  }{2
   H^{5/2}}\left(H_0-a t^2\right),
\end{equation}
where $b$ is a dimensionless constant. Its value $b\approx 12$ can be roughly estimated approximating the velocity profile by a parabola, as in the  gap-averaged Darcy's law \citep{shelley_hele_1997,bohr1994viscous,amar_fingering_2005,dias2013wavelength}. 

Approximation (\ref{eq:dpdrAcc}) is valid only for the cases of $\lambda \ll 1$ considered in this study. Since the ratio of the axial to the radial velocity, which can be estimated from (\ref{eq:Roft}), is comparable with $\lambda$, the effect of the axial velocity on the pressure gradient is negligibly small.

\subsection{Planar interface, long-wave approximation of small flow perturbations}

Since the radius of the liquid bridge is much larger than the gap thickness, $R \gg H$, the flow  leading to small interface disturbances can be considered in a Cartesian coordinate system $\{x, y\}$, where the $x$-coordinate coincides with the radial direction normal to the meniscus, defined as $x=0$, and the $y$-direction is tangential to the meniscus. The coordinate system $\{x, y\}$ is fixed at the meniscus of the liquid bridge, such that
\begin{equation}
 r = R(t) + x.
\end{equation}

The small flow perturbations in the direction normal to the substrates is neglected. Liquid flow occupies the half-infinite space $x\in ]-\infty;\;0]$. 
Denote $\{u'(x,y,t); \;v'(x,y,t)\}$ as the velocity vector of the flow perturbations, averaged through the gap width, and $p'(x,y,t)$ is the pressure perturbation. The absolute velocity and the pressure $p$ in the gap can be expressed in the form
\begin{equation}
u = \dot R(t) + u'(x,y,t),\quad v = v'(x,y,t),\quad p = p_0(x,t) + p'(x,y,t).
\end{equation}

Therefore, the time derivatives of the components of the velocity field can be written in the form
\begin{equation}
 u_{,t}= \ddot R + u'_{,t} - \dot R u'_{,x}, \quad
 v_{,t}= v'_{,t} - \dot R v'_{,x}.
\end{equation}

The gap thickness $H$ is assumed to be the smallest length scale in the problem. In this case consideration of only the dominant terms in the Navier-Stokes equation, written in the accelerating coordinate system, yields
\begin{subequations}
\begin{eqnarray}
 p_{,x} &=& \mu \left(-b \frac{\dot R+ u'}{H^2} +  u'_{,xx} + u'_{,yy}\right) - \rho \ddot R - \rho u'_{,t},\label{dpdx}\\
 p_{,y} &=& \mu \left(-b \frac{v'}{H^2} +  v'_{,xx} + v'_{,yy}\right) - \rho v'_{,t},\label{dpdy}
\end{eqnarray}
\end{subequations}
 
The characteristic value of the leading viscous terms in the pressure gradient expressions  (\ref{dpdx}) and (\ref{dpdy}) is $\mu b u'/H_0^2$. The characteristic time of the problem is $\sqrt{H_0/a}$. Therefore, the inertial terms of the flow fluctuations are of order $\rho u'_{,t}\sim \rho u' \sqrt{a/H_0}$. The Reynolds number, defined as the ratio of the inertial and viscous terms, is therefore
\begin{equation}\label{eq:Rey}
    Re = \frac{a^{1/2} H_0^{3/2}\rho}{b \mu}.
\end{equation}
In all our experiments the Reynolds number is of order $10^{-2}$. The inertial effects associated with the flow fluctuations are therefore negligibly small. 
The governing equation for the velocity perturbation can be then obtained from (\ref{dpdx}) and (\ref{dpdy}) neglecting the terms $\rho u'_{,t}$ and $\rho v'_{,t}$
\begin{equation}\label{prescond}
     -b \frac{u'_{,y}-v'_{,x}}{H^2} +  u'_{,xxy} + u'_{,yyy} - v'_{,xxx} - v'_{,yyx}=0.
\end{equation}
 
 The velocity field $\{u', v'\}$ has to satisfy (\ref{prescond}) as well as the continuity equation and the condition of the shear free meniscus surface
 \begin{subequations}
  \begin{eqnarray}\label{visc}
     u'_{,x}+v'_{,y}=0 ,\label{cont}\\ 
     u'_{,y}+v'_{,x}=0, \quad \mathrm{at}\quad x=0.\label{shearfree}
 \end{eqnarray}
 \end{subequations}
 
 Consider the sinusoidal profile of the flow fluctuations along the $y$ direction. This means that both velocity components include the term $\exp(i k y)$ where $k$ is the wavenumber.  The corresponding velocity field for the velocity of the small flow disturbances satisfying (\ref{prescond})-(\ref{shearfree}) is
\begin{subequations}
\begin{eqnarray}
     u' &=& \left( \exp(kx)-\frac{2H^2k^2}{b+2H^2k^2}\exp\left[\frac{\sqrt{b+H^2 k^2}}{H}x\right]\right)\exp(i k y ) T(t),\label{epru}\\
     v' &=&i\left( \exp(kx)-\frac{2H k \sqrt{b+H^2 k^2}}{b+2H^2k^2}\exp\left[\frac{\sqrt{b+H^2 k^2}}{H}x\right]\right)\exp(i k y ) T(t),
 \end{eqnarray}
 \end{subequations}
 where $T(t)$ is a function of time.

The small perturbations of the meniscus shape, defined as $x=\delta(y,t)$, are determined by the normal velocity component $u$ at the meniscus $x=0$. The boundary conditions for the meniscus perturbations, $\delta_{,t}=u$ at $x=0$, yield
\begin{equation}\label{eprdel}
\delta = \exp(i k y) G(t),\quad \mathrm{with}\quad  T(t) =\frac{b+2 H^2 k^2}{b} \dot G(t).
 \end{equation}

The pressure increment, associated with the flow perturbations at the interface, $p'$, is determined by the capillary forces and viscous stress 
 \begin{equation}\label{pvisc1}
p'(x,t) =- \sigma \delta_{,yy}-2 \mu u_{,x},\quad \mathrm{at}\quad x =\delta(y,t).
 \end{equation}

Note that the total pressure near the meniscus depends also on the curvature in the plane normal to the substrate. In this study the dependence of the shape of the meniscus in this plane on $\delta(y,t)$ is neglected, since the capillary pressure associated with this curvature is approximated by $p\sim \sigma/H$. Thus this pressure does not depend on the $y$-coordinate and does not contribute to the flow stability. 

The pressure $p'$ at the position $x=0$ can be approximated accounting for the smallness of the shape deformation
 \begin{equation}\label{pvisc}
p' =- \sigma \delta_{,yy} - \delta p_{0,r}-2 \mu u_{,x},\quad \mathrm{at}\quad x =0,
 \end{equation}
 where $p_{0}$ is the pressure gradient at the meniscus of the basic flow, determined in (\ref{eq:dpdrAcc}). The term $\delta p_{0,r}$ appears as a result of linearization of the pressure terms in the neighborhood of the liquid bridge  interface.
 
 Substituting (\ref{pvisc}) in expression (\ref{dpdy}) yields, with the help of (\ref{epru})-(\ref{eprdel}), the following ordinary differential equation for the function $G(t)$
 \begin{equation}\label{eqT}
b \left(k^3 \sigma -k p_{0,r}\right) H^2 G(t) +\left[4 H^3 k^3\left(\sqrt{H^2 k^2+b}-H k\right)
   +b^2\right] \mu  \dot G(t)=0
 \end{equation}

The solution of the ordinary differential equation (\ref{eqT}) is
  \begin{equation}\label{Tsol}
    G(t) = \delta_0 \exp\left[-\frac{b}{\mu}\int_0^t \frac{ H^2  \left(k^3 \sigma -k p_{0,r}\right)}{4 H^3 k^3\left(\sqrt{H^2 k^2+b}-H k\right)
   +b^2 }\mathrm{d}t\right],
 \end{equation}
 where $\delta_0$ is the initial meniscus perturbation.

 The function $G(t)$ in (\ref{Tsol}) can be derived using (\ref{eq:Roft}) and (\ref{eq:dpdrAcc}). It can be expressed in dimensionless form
 \begin{eqnarray}
     G&=&\delta_0\exp\left[ \frac{\sqrt{b}}{2\lambda}\;\int_0^\tau\Omega(\xi,\tau)\mathrm{d}\tau\right],\label{ampl}\\
     \Omega&=&\frac{  \tau \xi \left(\tau ^2+1\right)^{-3/2} -\frac{\xi
   ^3}{Ca}{\left(\tau ^2+1\right)^{2}} + \frac{Re \left(1-2 \tau^2\right) \xi }{\sqrt{2} \sqrt{\tau^2+1}} }{1-4 \left(\tau ^2+1\right)^3 \xi ^3 \left[\xi  (\tau ^2+1) -\sqrt{(\tau
   ^2+1)^2 \xi ^2+1}\right]},\label{omegadef}
 \end{eqnarray}
 where the dimensionless time $\tau$, the dimensionless wavenumber $\xi$ and the capillary number $Ca$ are defined as
   \begin{equation}\label{parms}
\tau = t\sqrt{\frac{a}{2 H_0}}, \quad \xi = \frac{k H_0}{\sqrt{b}},\quad Ca=\frac{\sqrt{a} \mu  R_0}{\sqrt{2 H_0}  \sigma }.
 \end{equation}
The Reynolds number is defined in (\ref{eq:Rey}) and the geometrical parameter is $\lambda = H_0/2 R_0$, as defined in \S\ref{sec:exp}. 

Equations (\ref{ampl}) and (\ref{omegadef}) allow computation of the evolution of the amplitude of the waves for a given wavelength and given parameters of the liquid bridge stretching. 

\begin{figure}
    \centering
    \setlength{\figW}{12. cm}%
    \setlength{\figH}{9 cm}%
\input{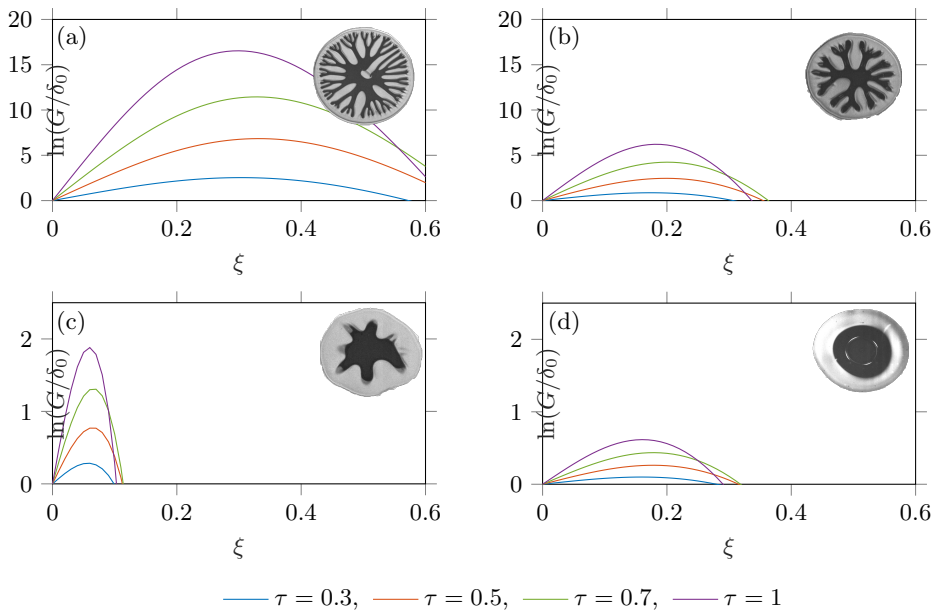}
\caption{Dimensionless  amplitude of the radius perturbations $\ln(G/\delta_0)$ as a function of the dimensionless wavenumbers $\xi$ for various time instants $\tau$, computed by numerical integration of (\ref{Tsol}). (a) $Ca=2.45, Re=0.005, \lambda=0.0059$, (b) $Ca=0.704, Re=0.0026, \lambda=0.0099$, (c) $Ca=0.0663, Re=0.0252, \lambda=0.0109$ and (d) $Ca=0.5109, Re=0.034, \lambda=0.0902$. }
\label{fig:functionG}
\end{figure}

In fig.~\ref{fig:functionG} the dimensionless amplitude of the perturbations of the bridge radius $\ln(G/\delta_0)$, computed using (\ref{ampl}), is shown as a function of the dimensionless wavenumber $\xi$ for various times $\tau$. In the cases shown in fig.~\ref{fig:functionG}a) and b), corresponding to a large number of fingers, the absolute amplitude of the perturbation $G$ is two orders of magnitude higher than the amplitude of the initial perturbations $\delta_0$. In the case fig.~\ref{fig:functionG}c) close to the fingering threshold, the value of $G/\delta_0\sim 10^1$, while in the case shown in fig.~\ref{fig:functionG}d) in which no apparent fingering has been observed, the value of $G/\delta_0$ is of order unity.
In each of the cases shown in fig.~\ref{fig:functionG}a) the wavenumber corresponding to the maximum amplitude is only slightly dependent on  time, but is significantly influenced by the parameters of the bridge stretching.

\subsection{Approximation for small capillary numbers}
In the long-wave approximation, values of $\xi$ are assumed to be small. This assumption can again be examined after the solution for typical values of $\xi$ has been obtained. In this study only the dominant terms are taken into account, while the terms of order $O(\xi^4)$ are neglected. The corresponding approximated expression for $\int_0^\tau\Omega\mathrm{d}\tau$ is derived in the form
\begin{eqnarray}
    \int_0^\tau\Omega(\xi,\tau)\mathrm{d}\tau &=&\xi-\frac{\tau  \left(3 \tau ^4+10 \tau ^2+15\right) \xi ^3}{15 Ca}-\frac{\xi }{\sqrt{\tau ^2+1}}\nonumber\\
    &-&\frac{Re \xi}{\sqrt{2}}\left(\tau  \sqrt{\tau ^2+1}-2 \mathrm{arcsinh}\tau \right)+\mathcal{O}(\xi^4).\label{eqOmeg}
\end{eqnarray}

The most unstable mode $\xi_\ast$ associated with the maximum positive value of the function $\int_0^\tau\Omega\mathrm{d}\tau$ is therefore
  \begin{equation}\label{chistar}
   \xi_\ast =\sqrt{Ca}\left[\frac{1-\frac{1}{\sqrt{\tau ^2+1}}+ \frac{Re}{\sqrt{2}}\left(\tau  \sqrt{\tau ^2+1}-2 \mathrm{arcsinh}\tau \right)}{\tau  \left(\frac{3}{5} \tau ^4+2 \tau ^2+3\right)}\right]^{1/2}.
 \end{equation}

 The dimensionless time $\tau$ is of order of  unity. The value of the dimensionless wavenumber has to also be  small in the framework of the long-wave approximation used in this study. Therefore, the solution (\ref{chistar}) for the most unstable mode $\xi_*$ is valid only for small capillary numbers, as claimed in the title of this section.

 The wavelength of the most unstable mode is $\ell_\ast=2\pi/k$. The number of finger-like jets is therefore 
 \begin{equation}\label{Nfmax}
     N_f=2\pi R/\ell_\ast= \frac{\sqrt{b}}{2\lambda} \frac{\xi_\ast}{\sqrt{1+\tau^2}}.
 \end{equation}

The expression for the number of fingers is obtained with the help of (\ref{chistar})
   \begin{equation}\label{Nstar}
N_f = \frac{\sqrt{b Ca}}{2\lambda}\left[\frac{1-\frac{1}{\sqrt{\tau ^2+1}}+ \frac{Re}{\sqrt{2}}\left(\tau  \sqrt{\tau ^2+1}-2 \mathrm{arcsinh}\tau \right)}{\tau (\tau ^2+1) \left(\frac{3}{5} \tau ^4+2 \tau ^2+3\right)}\right]^{1/2}.
 \end{equation}

 The predicted number of the fingers depends on the dimensionless time $\tau$. Such dependence is confirmed by observations.

\section{Results and Discussion}

The maximum value of the function  $N_f(\tau)$  can be computed from  (\ref{Nstar}).  In the limit $Re=0$ the maximum 
\begin{equation}\label{eq:NmaxSmallCa}
    N_{\mathrm{max}}\approx 0.38 \sqrt{Ca}/\lambda,\quad Ca \ll 1,\quad Re =0.
\end{equation}
is reached at the instant $\tau_\star\approx 0.49$. The predicted bridge radius $R$ corresponding to the maximum number of jets is therefore $R_\star=R_0(1+\tau_\star^2)^{-1/2}\approx 0.9 R_0$. 

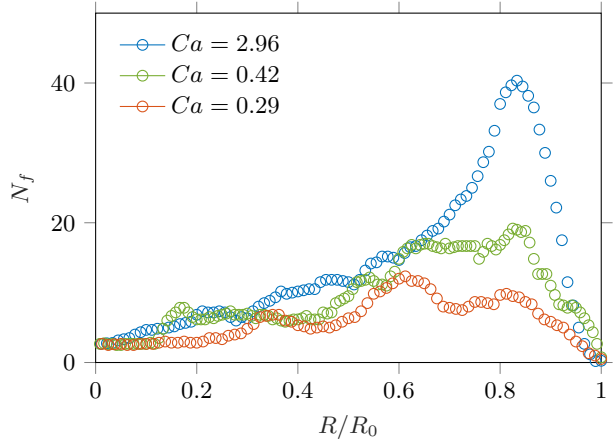
\begin{figure}
    \centering
    \setlength{\figW}{7. cm}%
    \setlength{\figH}{4.6 cm}%
%
%
\definecolor{mycolor1}{rgb}{0.00000,0.44700,0.74100}%
\definecolor{mycolor2}{rgb}{0.85000,0.32500,0.09800}%
\definecolor{mycolor3}{rgb}{0.46600,0.67400,0.18800}%
\definecolor{mycolor4}{rgb}{0.4940, 0.1840, 0.5560}%

\begin{tikzpicture}

\begin{axis}[%
width=0.955\figW,
height=\figH,
at={(0\figW,0\figH)},
scale only axis,
xmin=0,
xmax=1,
xlabel style={font=\color{white!15!black}},
xlabel={$R/R_{0}$},
ymin=0,
ymax=50,
ylabel style={font=\color{white!15!black}},
ylabel={$N_f$},
tick align=outside,
axis background/.style={fill=white},
legend style={at={(0.03,0.97)}, anchor=north west, legend cell align=left, align=left, fill=none, draw=none}
]
\addplot [color=mycolor1, draw=none, mark=o, mark options={solid, mycolor1}]
  table[row sep=crcr]{%
0.0111111111111111	2.66666666666667\\
0.0222222222222222	2.75\\
0.0333333333333333	2.8\\
0.0444444444444444	3.16666666666667\\
0.0555555555555556	3.33333333333333\\
0.0666666666666667	3.5\\
0.0777777777777778	4\\
0.0888888888888889	4.33333333333333\\
0.1	4.66666666666667\\
0.111111111111111	4.66666666666667\\
0.122222222222222	4.83333333333333\\
0.133333333333333	4.83333333333333\\
0.144444444444444	5\\
0.155555555555556	5.16666666666667\\
0.166666666666667	5.5\\
0.177777777777778	5.66666666666667\\
0.188888888888889	5.83333333333333\\
0.2	6.83333333333333\\
0.211111111111111	7.33333333333333\\
0.222222222222222	7.16666666666667\\
0.233333333333333	7.16666666666667\\
0.244444444444444	7.33333333333333\\
0.255555555555556	7.16666666666667\\
0.266666666666667	6.5\\
0.277777777777778	6\\
0.288888888888889	6.5\\
0.3	6.33333333333333\\
0.311111111111111	7.33333333333333\\
0.322222222222222	8\\
0.333333333333333	8.5\\
0.344444444444444	8.83333333333333\\
0.355555555555556	9.33333333333333\\
0.366666666666667	10.1666666666667\\
0.377777777777778	9.83333333333333\\
0.388888888888889	10\\
0.4	10.1666666666667\\
0.411111111111111	10.3333333333333\\
0.422222222222222	10.5\\
0.433333333333333	10.5\\
0.444444444444444	11.5\\
0.455555555555556	11.8333333333333\\
0.466666666666667	11.8333333333333\\
0.477777777777778	11.8333333333333\\
0.488888888888889	11.6666666666667\\
0.5	11.5\\
0.511111111111111	11.1666666666667\\
0.522222222222222	11.8333333333333\\
0.533333333333333	13.1666666666667\\
0.544444444444445	14.1666666666667\\
0.555555555555556	14.3333333333333\\
0.566666666666667	15.1666666666667\\
0.577777777777778	15.1666666666667\\
0.588888888888889	15\\
0.6	14.6666666666667\\
0.611111111111111	15.6666666666667\\
0.622222222222222	16.5\\
0.633333333333333	16.6666666666667\\
0.644444444444444	17.5\\
0.655555555555556	18.1666666666667\\
0.666666666666667	18.8333333333333\\
0.677777777777778	19.1666666666667\\
0.688888888888889	20.1666666666667\\
0.7	21.1666666666667\\
0.711111111111111	22.5\\
0.722222222222222	23.3333333333333\\
0.733333333333333	23.8333333333333\\
0.744444444444444	25\\
0.755555555555556	26.6666666666667\\
0.766666666666667	28.6666666666667\\
0.777777777777778	30.1666666666667\\
0.788888888888889	33.1666666666667\\
0.8	37\\
0.811111111111111	38.6666666666667\\
0.822222222222222	39.6666666666667\\
0.833333333333333	40.3333333333333\\
0.844444444444444	39.5\\
0.855555555555556	38.1666666666667\\
0.866666666666667	36.5\\
0.877777777777778	33.3333333333333\\
0.888888888888889	30\\
0.9	26\\
0.911111111111111	22.1666666666667\\
0.922222222222222	17.5\\
0.933333333333333	11.5\\
0.944444444444444	8.16666666666667\\
0.955555555555556	5\\
0.966666666666667	2.66666666666667\\
0.977777777777778	1.16666666666667\\
0.988888888888889	0.2\\
1	0.25\\
};
\addlegendentry{$Ca=2.96$}

\addplot [color=mycolor3, draw=none, mark=o, mark options={solid, mycolor3}]
  table[row sep=crcr]{%
0.00833333333333333	2.66666666666667\\
0.0166666666666667	2.75\\
0.025	2.6\\
0.0333333333333333	2.66666666666667\\
0.0416666666666667	2.5\\
0.05	2.5\\
0.0583333333333333	2.66666666666667\\
0.0666666666666667	2.66666666666667\\
0.075	2.66666666666667\\
0.0833333333333333	2.66666666666667\\
0.0916666666666667	2.66666666666667\\
0.1	2.66666666666667\\
0.108333333333333	2.83333333333333\\
0.116666666666667	2.66666666666667\\
0.125	3.5\\
0.133333333333333	4.83333333333333\\
0.141666666666667	6\\
0.15	6.66666666666667\\
0.158333333333333	7.16666666666667\\
0.166666666666667	7.83333333333333\\
0.175	7.83333333333333\\
0.183333333333333	7.16666666666667\\
0.191666666666667	6.5\\
0.2	6.16666666666667\\
0.208333333333333	6.16666666666667\\
0.216666666666667	6.16666666666667\\
0.225	6.5\\
0.233333333333333	6.33333333333333\\
0.241666666666667	6.66666666666667\\
0.25	6.83333333333333\\
0.258333333333333	7\\
0.266666666666667	7.33333333333333\\
0.275	6.83333333333333\\
0.283333333333333	6.83333333333333\\
0.291666666666667	6.83333333333333\\
0.3	7\\
0.308333333333333	6.66666666666667\\
0.316666666666667	6.33333333333333\\
0.325	6.83333333333333\\
0.333333333333333	6.83333333333333\\
0.341666666666667	6.66666666666667\\
0.35	6.16666666666667\\
0.358333333333333	6.33333333333333\\
0.366666666666667	6\\
0.375	5.83333333333333\\
0.383333333333333	6\\
0.391666666666667	6.16666666666667\\
0.4	6.16666666666667\\
0.408333333333333	6.33333333333333\\
0.416666666666667	6.66666666666667\\
0.425	6.16666666666667\\
0.433333333333333	6.16666666666667\\
0.441666666666667	6.16666666666667\\
0.45	7.16666666666667\\
0.458333333333333	7\\
0.466666666666667	7.16666666666667\\
0.475	8.5\\
0.483333333333333	8.83333333333333\\
0.491666666666667	9.16666666666667\\
0.5	9.5\\
0.508333333333333	10.5\\
0.516666666666667	11.5\\
0.525	11.6666666666667\\
0.533333333333333	11.8333333333333\\
0.541666666666667	12.1666666666667\\
0.55	11.6666666666667\\
0.558333333333333	11.1666666666667\\
0.566666666666667	11.1666666666667\\
0.575	11\\
0.583333333333333	12.3333333333333\\
0.591666666666667	13\\
0.6	14.8333333333333\\
0.608333333333333	15.8333333333333\\
0.616666666666667	16.3333333333333\\
0.625	16.8333333333333\\
0.633333333333333	16.3333333333333\\
0.641666666666667	16.8333333333333\\
0.65	17\\
0.658333333333333	17\\
0.666666666666667	16.6666666666667\\
0.675	16.5\\
0.683333333333333	16.3333333333333\\
0.691666666666667	17\\
0.7	16.3333333333333\\
0.708333333333333	16.5\\
0.716666666666667	16.6666666666667\\
0.725	16.6666666666667\\
0.733333333333333	16.6666666666667\\
0.741666666666667	16.6666666666667\\
0.75	16.6666666666667\\
0.758333333333333	14.8333333333333\\
0.766666666666667	15.8333333333333\\
0.775	17\\
0.783333333333333	16.6666666666667\\
0.791666666666667	16.1666666666667\\
0.8	16.8333333333333\\
0.808333333333333	17.5\\
0.816666666666667	18.3333333333333\\
0.825	19.1666666666667\\
0.833333333333333	19\\
0.841666666666667	18.6666666666667\\
0.85	18.8333333333333\\
0.858333333333333	17.1666666666667\\
0.866666666666667	14.8333333333333\\
0.875	12.6666666666667\\
0.883333333333333	12.6666666666667\\
0.891666666666667	12.6666666666667\\
0.9	11\\
0.908333333333333	9.66666666666667\\
0.916666666666667	8.33333333333333\\
0.925	8.16666666666667\\
0.933333333333333	7.83333333333333\\
0.941666666666667	8.16666666666667\\
0.95	7.83333333333333\\
0.958333333333333	7.16666666666667\\
0.966666666666667	6.33333333333333\\
0.975	5.33333333333333\\
0.983333333333333	4.33333333333333\\
0.991666666666667	2.66666666666667\\
1	0.75\\
};
\addlegendentry{$Ca=0.42$}

\addplot [color=mycolor2, draw=none, mark=o, mark options={solid, mycolor2}]
  table[row sep=crcr]{%
0.0125	2.66666666666667\\
0.025	2.5\\
0.0375	2.6\\
0.05	2.66666666666667\\
0.0625	2.5\\
0.075	2.5\\
0.0875	2.66666666666667\\
0.1	2.66666666666667\\
0.1125	2.66666666666667\\
0.125	2.83333333333333\\
0.1375	3\\
0.15	2.83333333333333\\
0.1625	2.83333333333333\\
0.175	3\\
0.1875	2.83333333333333\\
0.2 2.83333333333333\\
0.2125	3\\
0.225	3.5\\
0.2375	3.5\\
0.25	3.33333333333333\\
0.2625	3.83333333333333\\
0.275	4\\
0.2875	4.16666666666667\\
0.3	4.83333333333333\\
0.3125	5.5\\
0.325	6.5\\
0.3375	6.66666666666667\\
0.35	6.83333333333333\\
0.3625	6.83333333333333\\
0.375	6\\
0.3875	5.83333333333333\\
0.4	5.33333333333333\\
0.4125	5.16666666666667\\
0.425	4.83333333333333\\
0.4375	5\\
0.45	5.16666666666667\\
0.4625	5\\
0.475	5.33333333333333\\
0.4875	5.83333333333333\\
0.5	6.33333333333333\\
0.5125	6.5\\
0.525	7.33333333333333\\
0.5375	8.5\\
0.55	9.83333333333333\\
0.5625	10.1666666666667\\
0.575	10.6666666666667\\
0.5875	11.3333333333333\\
0.6	12\\
0.6125	12.3333333333333\\
0.625	11.6666666666667\\
0.6375	11.6666666666667\\
0.65	11.3333333333333\\
0.6625	10.5\\
0.675	9.16666666666667\\
0.6875	8\\
0.7	7.83333333333333\\
0.7125	7.66666666666667\\
0.725	7.5\\
0.7375	8\\
0.75	8.66666666666667\\
0.7625	8.66666666666667\\
0.775	8.5\\
0.7875	8.33333333333333\\
0.8	9.66666666666667\\
0.8125	9.83333333333333\\
0.825	9.5\\
0.8375	9.33333333333333\\
0.85	8.66666666666667\\
0.8625	8.33333333333333\\
0.875	7\\
0.8875	6.16666666666667\\
0.9	5.83333333333333\\
0.9125	5.33333333333333\\
0.925	5\\
0.9375	4\\
0.95	3\\
0.9625	2.33333333333333\\
0.975	1.33333333333333\\
0.9875	1\\
1	0.5\\
};
\addlegendentry{$Ca=0.29$}

\end{axis}
\end{tikzpicture}%
\caption{The  number of fingers $N_f$ as a function of the liquid bridge radius $R$, observed in three different experiments. The  measurements were performed with $\lambda \approx 0.01$ and $Re \approx 0.1$.}
\label{fig:functionRR0}
\end{figure}

In fig.~\ref{fig:functionRR0} The experimentally measured number of fingers at various  instants and corresponding radii scaled by $R_0$ are shown exemplary for three different values of $Ca$ but for nearly same values of $\lambda$ and $Re$. The number of fingers reaches the maximum at the radii $R_\star/R_0 \approx 0.8 -0.9$, as predicted by the theory.

For lower radii $R\ll R_\star$, corresponding to longer times $\tau$, the flow in the gap is significantly influenced by the non-linear effects associated with the growth of fingers. Such non-linear analysis is out of the scope of this theoretical study. For smaller $Ca$ numbers the influence of the non-linear effects gets larger, for those cases we have to limit our analysis to local maximum at $R>R_\star$.

The amplitude of the perturbations at the corresponding conditions, $\xi=\xi_\star$, $\tau=\tau_\star$, can also be estimated for small capillary numbers
\begin{equation}\label{Gstar}
    G_\star \approx \delta_0 \exp(0.11 Ca^{1/2} \lambda^{-1}),\quad Ca \ll 1,\quad Re =0.
\end{equation}

\begin{figure}
    \centering
    \setlength{\figW}{7. cm}%
    \setlength{\figH}{4.6 cm}%
%
%

\definecolor{mycolor1}{rgb}{0.00000,0.44700,0.74100}%
\definecolor{mycolor2}{rgb}{0.85000,0.32500,0.09800}%
\definecolor{mycolor3}{rgb}{0.46600,0.67400,0.18800}%
\pgfplotsset{every  tick/.style={black,}, minor x tick num=1,}
\begin{tikzpicture}

\begin{axis}[%
width=0.955\figW,
height=\figH,
at={(0\figW,0\figH)},
scale only axis,
minor x tick num=3,
minor y tick num=3,
xtick={0, 0.04, 0.08, 0.12, 0.16, 0.2},ytick={0, 0.25, 0.5, 0.75},
xmin=0,
xmax=0.2,
xlabel style={font=\color{white!15!black}},
xlabel={$Re$},
ymin=0,
ymax=0.75,
ylabel style={font=\color{white!15!black}},
ylabel={$N_{\mathrm{max}} \lambda Ca^{-1/2}$},
tick align=outside,
axis background/.style={fill=white},
legend style={legend cell align=left, align=left, fill=none, draw=none}
]
\addplot [only marks, color=mycolor1, draw=none, mark=diamond, mark options={solid, scale=1.5, fill=mycolor1, mycolor1}]
  table[row sep=crcr]{%
0.0347986104374134	0.266461050557045\\
0.0401589587121458	0.295058821849393\\
0.00238081398843284	0.247479866095415\\
0.0103583152561806	0.241451737369453\\
0.0210140336817418	0.387488359167826\\
0.0502766642940994	0.461137462790065\\
0.00264992670847276	0.283279996722654\\
0.00586124945711548	0.364734308083737\\
0.0109675108722367	0.264485151661834\\
0.00255804400791482	0.318129082489562\\
0.0508507254384433	0.447408170205787\\
0.00262073677458277	0.211485534516622\\
0.0988076316419428	0.24641646161473\\
0.0060492577730548	0.353423483713361\\
0.00282291135000305	0.209460942655003\\
0.105009562234056	0.303217362691136\\
0.0251699998887845	0.4647516086263\\
0.108092735197654	0.25220917136121\\
0.00318717960523582	0.410894661371856\\
0.0309623094087589	0.338216056830981\\
0.0681158129160981	0.306545660566711\\
0.0145237973394626	0.334144932366071\\
0.0312023079054439	0.251844326413687\\
0.00974551660047577	0.455070049149106\\
};
\addlegendentry{Experiments}

\addplot [color=black, dashed]
  table[row sep=crcr]{%
0	0.38\\
0.01	0.373\\
0.02	0.368\\
0.03	0.363\\
0.04	0.358\\
0.06	0.348\\
0.08	0.338\\
0.1	0.329\\
0.15	0.306\\
0.2	0.286\\
0.3	0.252251\\
};
\addlegendentry{Theory for $Ca<1$}

\end{axis}
\end{tikzpicture}%
\caption{Scaled maximum number of the observed fingers $N_\mathrm{max} \lambda/Ca^{1/2}$ as a function of the Reynolds number.} \label{fig:Nmax_scaled}
\end{figure}
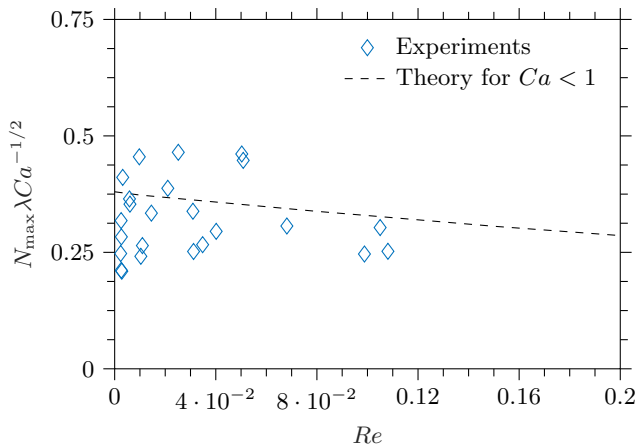

Following the results of the approximate estimation of the number of fingers (\ref{Nstar}), the dimensionless parameter $N_\mathrm{max} \lambda/Ca^{1/2}$ is a function of the Reynolds number if the capillary number is small. In fig.~\ref{fig:Nmax_scaled} this dependence is compared with theoretical predictions based on the numerical computation of the maximum value of the expression for $N_f$  (\ref{Nstar}).  The theoretical predictions do not contradict the experiments although the clear dependence of the number of fingers on the value of the Reynolds number is not so apparent because of the relatively large scatter of the data. This scatter is explained by the fact that in the cases close to the fingering threshold, the amplitude of perturbations is relatively small at the time instant corresponding to $N_f = N_\mathrm{max}$. The fingers therefore can only be recognized by our optical system slightly later, when the amplitude magnification is significant. The conditions near the fingering threshold (\ref{eq:lam_thr}) are discussed later in this section. 

\begin{figure}
    \centering
\includegraphics[width=.6\textwidth]{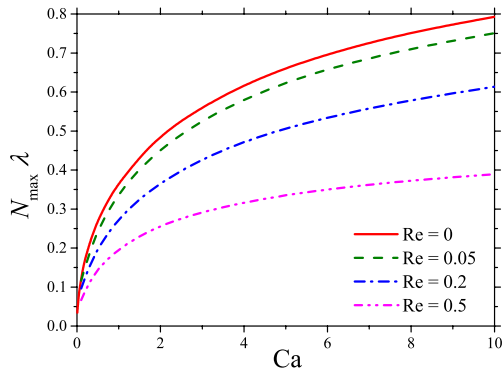}
\caption{Computational results of $N_\mathrm{max} \lambda$ as a function of the capillary number $Ca$ for various Reynolds numbers $Re$. Comparison with theoretical predictions based on the approximate solution.} \label{fig:Nmax_CaRe}
\end{figure}

 The approximate solution (\ref{chistar}) for the most unstable mode, based on the assumption of the smallness of $\xi_\ast$, is not applicable to the cases when the capillary number is not very small. In these cases a complete numerical solution is required. In this solution the values of $\xi_\ast(\tau)$ for a specific capillary number $Ca$ and Reynolds number are first computed as a point corresponding to the maximum  of $\int_0^\tau \Omega\mathrm{d}\tau$, where $\Omega(\xi,\tau)$ is defined in (\ref{omegadef}).  Then, the maximum number of fingers is computed with the help of (\ref{Nfmax}) at the time interval $\tau >0$. The theoretically predicted values of $N_\mathrm{max} \lambda$ are determined only by the capillary number and by the Reynolds number. The theoretical predictions of $N_\mathrm{max} \lambda$ are shown in fig.~\ref{fig:Nmax_CaRe}. As expected, the influence of inertia becomes significant when both the capillary and Reynolds numbers are relatively large. 
 
  \begin{figure}
    \centering
\includegraphics[width=.6\textwidth]{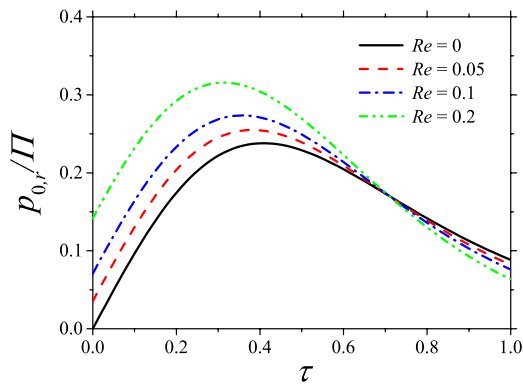}
\caption{The values of the scaled pressure gradient at the meniscus interface, $p_{0, r}/\Pi$ as a function of dimensionless time $\tau$ for various values of the Reynolds number $Re$. The scale for the pressure gradient, $\Pi$, is defined in (\ref{accPi}).} \label{fig:EffectA}
\end{figure}

The significance of the inertial effects in this problem is rather surprising, noting very small values of the Reynolds numbers considered in this study. The main factor governing the fingering process is the pressure gradient at the meniscus interface (\ref{eq:dpdrAcc}), obtained from the base solution. The mechanism of instability caused by the positive normal pressure gradient at the liquid interface is analogous to the Rayleigh-Taylor instability \citep{chandrasekhar2013hydrodynamic}, where this gradient is caused by gravity or interface acceleration. In our case this term can be written in the dimensionless form with the help of (\ref{eq:Rey}) and (\ref{parms})
\begin{equation}\label{accPi}
    \frac{p_{0,r}}{\Pi} =\frac{\tau }{\left(\tau ^2+1\right)^{7/2}} + \frac{\text{Re} \left(1-2 \tau ^2\right)}{\sqrt{2} \left(\tau
   ^2+1\right)^{5/2}} , \quad \Pi = \frac{\sqrt{a} b \mu }{2 \sqrt{2} H_0^{3/2} \lambda }.
\end{equation}

Function $p_{0,r}(\tau)/\Pi$ is shown in the plots in fig.~\ref{fig:EffectA} for various values of the Reynolds number. The inertial effects, associated with terms in (\ref{accPi}) including Reynolds number, is most pronounced at the very initial stages of the bridge stretching when the substrate velocity (and thus the viscous stresses) is small. This is why in the case of liquid bridge stretching by an accelerating substrate both viscous and inertial effects contribute to the meniscus instability. 

\begin{figure}
    \centering
    \setlength{\figW}{7. cm}%
    \setlength{\figH}{6 cm}%
%
%
\definecolor{mycolor1}{rgb}{0.00000,0.44700,0.74100}%
\definecolor{mycolor2}{rgb}{0.85000,0.32500,0.09800}%
\definecolor{mycolor3}{rgb}{0.46600,0.67400,0.18800}%
\pgfplotsset{every  tick/.style={black,}, minor x tick num=1,}

\begin{tikzpicture}

\begin{axis}[%
width=0.955\figW,
height=\figH,
at={(0\figW,0\figH)},
scale only axis,
minor x tick num=3,
minor y tick num=3,
xtick={0,40,80,120},ytick={0,40,80,120},
xmin=0,
xmax=120,
xlabel style={font=\color{white!15!black}},
xlabel={$N_{\mathrm{max}}$, theory},
ymin=0,
ymax=120,
ylabel style={font=\color{white!15!black}},
ylabel={$N_{\mathrm{max}}$, experiment},
tick align=outside,
axis background/.style={fill=white},
legend style={at={(0.03,0.97)}, anchor=north west, legend cell align=left, align=left, fill=none, draw=none}
]
\addplot [only marks, color=mycolor1, draw=none, mark=diamond, mark options={solid, scale=1.5, fill=mycolor1, mycolor1}]
  table[row sep=crcr]{%
35.8564	26\\
48.4421	40\\
46.4514	34\\
27.9254	28\\
15.24342	19\\
33.2692	34\\
31.6676	27\\
24.2859	30\\
32.9101	22\\
41.861	42\\
18.29586	15\\
8.74041	11\\
37.8187	29\\
17.96407	20\\
10.94284	10\\
24.23869	20\\
33.3796	32\\
35.4989	28\\
13.83211	17\\
};
\addlegendentry{without cavitation}

\addplot [only marks, color=mycolor3, draw=none, mark=square, mark options={solid, scale=1, fill=mycolor3, mycolor3}]
  table[row sep=crcr]{%
58.5411	40\\
 55.8407	46\\
 73.0878	43\\
 68.8841	40\\
};
\addlegendentry{transient cavitation}

\addplot [only marks, color=mycolor2, draw=none, mark=otimes*, mark options={solid, fill=mycolor2, mycolor2}]
  table[row sep=crcr]{%
95	50\\
102.026	54\\
88.1788	56\\
107.76	47\\
75.3264	59\\
95	51\\
107.94	52\\
};
\addlegendentry{with cavitation}

\addplot [color=black, dashed, forget plot]
  table[row sep=crcr]{%
0	0\\
1	1\\
2	2\\
3	3\\
4	4\\
5	5\\
6	6\\
7	7\\
8	8\\
9	9\\
10	10\\
11	11\\
12	12\\
13	13\\
14	14\\
15	15\\
16	16\\
17	17\\
18	18\\
19	19\\
20	20\\
21	21\\
22	22\\
23	23\\
24	24\\
25	25\\
26	26\\
27	27\\
28	28\\
29	29\\
30	30\\
31	31\\
32	32\\
33	33\\
34	34\\
35	35\\
36	36\\
37	37\\
38	38\\
39	39\\
40	40\\
41	41\\
42	42\\
43	43\\
44	44\\
45	45\\
46	46\\
47	47\\
48	48\\
49	49\\
50	50\\
51	51\\
52	52\\
53	53\\
54	54\\
55	55\\
56	56\\
57	57\\
58	58\\
59	59\\
60	60\\
61	61\\
62	62\\
63	63\\
64	64\\
65	65\\
66	66\\
67	67\\
68	68\\
69	69\\
70	70\\
71	71\\
72	72\\
73	73\\
74	74\\
75	75\\
76	76\\
77	77\\
78	78\\
79	79\\
80	80\\
81	81\\
82	82\\
83	83\\
84	84\\
85	85\\
86	86\\
87	87\\
88	88\\
89	89\\
90	90\\
91	91\\
92	92\\
93	93\\
94	94\\
95	95\\
96	96\\
97	97\\
98	98\\
99	99\\
100	100\\
101	101\\
102	102\\
103	103\\
104	104\\
105	105\\
106	106\\
107	107\\
108	108\\
109	109\\
110	110\\
111	111\\
112	112\\
113	113\\
114	114\\
115	115\\
116	116\\
117	117\\
118	118\\
119	119\\
120	120\\
121	121\\
122	122\\
123	123\\
124	124\\
125	125\\
126	126\\
127	127\\
128	128\\
129	129\\
130	130\\
131	131\\
132	132\\
133	133\\
134	134\\
135	135\\
136	136\\
137	137\\
138	138\\
139	139\\
140	140\\
};
\end{axis}
\end{tikzpicture}%
\caption{Comparison of the measured and theoretically predicted maximum number of fingers $N_\mathrm{max}$. The experiments accompanied by cavitation are marked by circles. The straight dashed line corresponds to perfect agreement between experiment and theory. } \label{fig:Nmax_theory}
\end{figure}
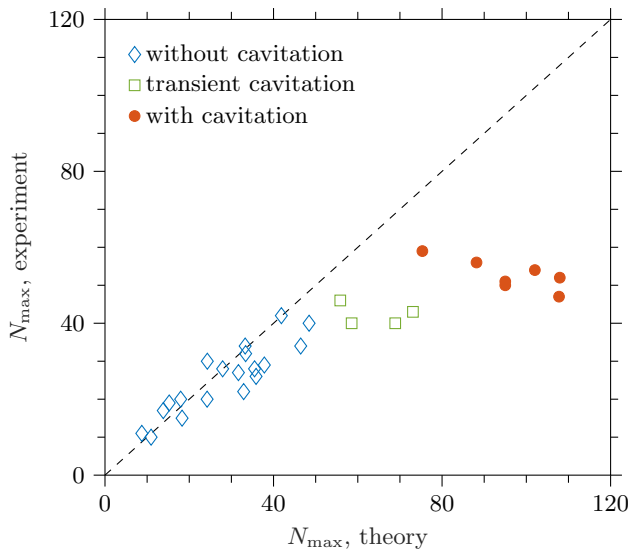
 
 The experimental and theoretically predicted values for $N_\mathrm{max}$ are compared in fig.~\ref{fig:Nmax_theory}. The agreement is rather good for most of the cases. For some cases, however, the number of fingers is overestimated. 

\begin{figure}
    \centering
\includegraphics[width=.6\textwidth]{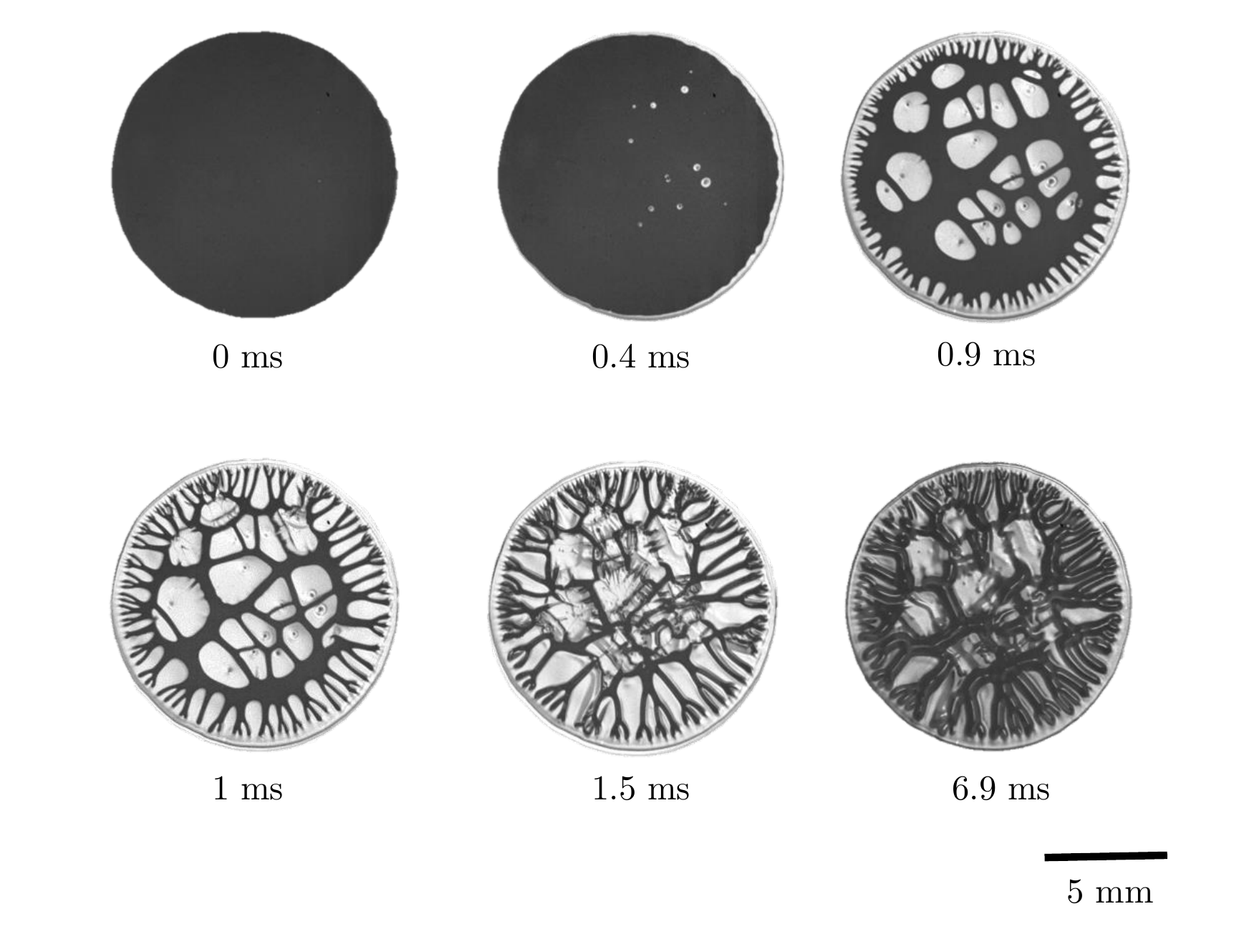}
\caption{Example of the void formation during liquid bridge stretching. The liquid is \textit{Gly80}. The other experimental parameters are: $a=\SI{180}{m/s^2}$, $H_0 =\SI{60}{\mu m}$, $\lambda= 0.006$. } \label{fig:cavitation}
\end{figure}

In all these overestimated experiments several voids in the liquid bridge have been observed, formed due to cavitation.  In some cases these voids quickly expand, leading to the formation of the structures resembling  the Voronoi tessellation, as shown in the examples in fig.~\ref{fig:cavitation}. These cases are marked as liquid bridge stretching \emph{with cavitation}. 

\begin{figure}
    \centering
\includegraphics[width=.6\textwidth]{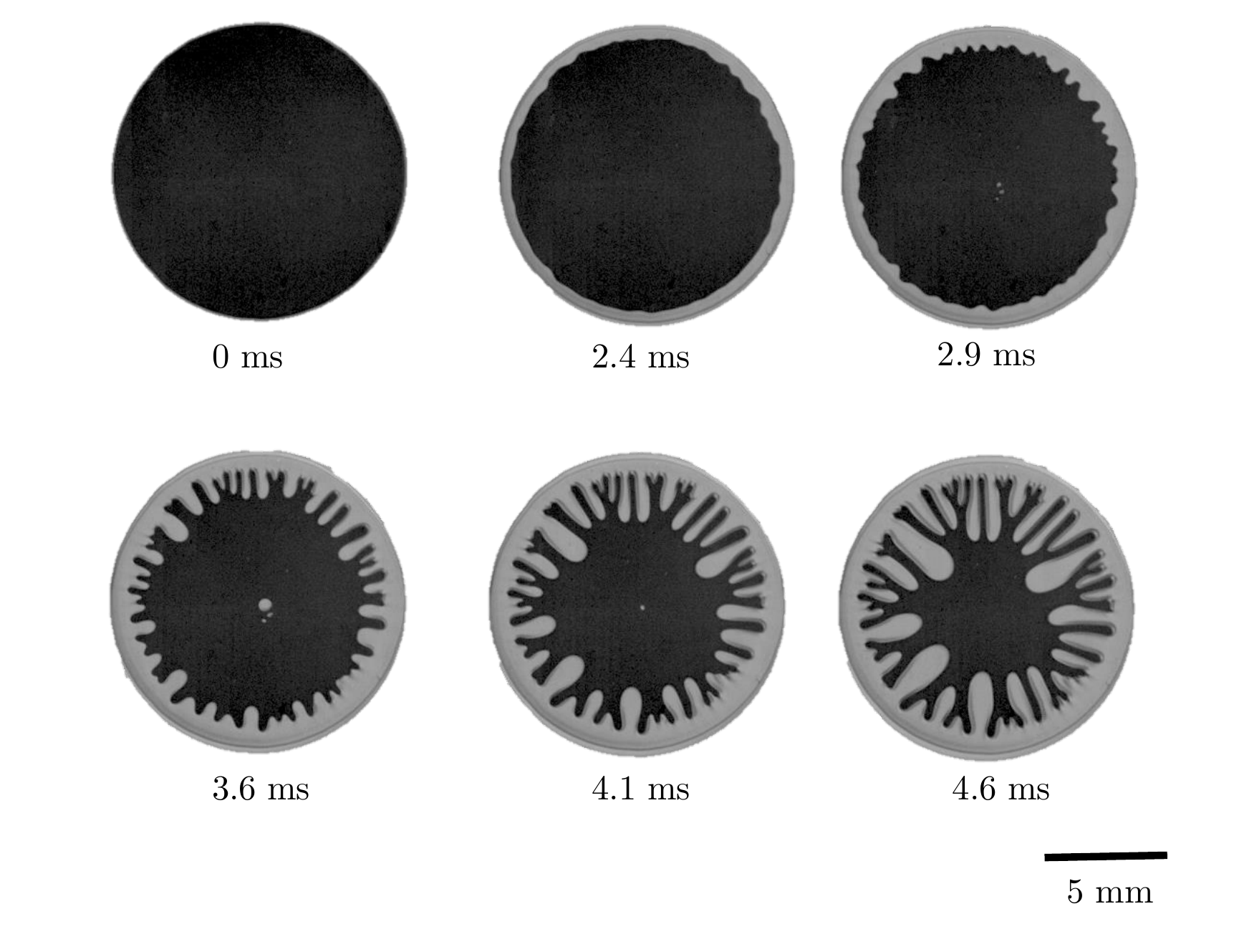}
\caption{Example of the transient cavitation. Several voids are formed in the central part of the liquid bridge and then disappear. The liquid is \textit{Gly80}. The other experimental parameters are: $a=\SI{10}{m/s^2}$, $H_0 =\SI{53}{\mu m}$, $\lambda= 0.006$. } \label{fig:cavitation_transient}
\end{figure}

Several additional cases have also been observed, marked in fig. \ref{fig:cavitation_transient} as \emph{transient cavitation}. In these cases a small number of macroscopic voids emerge far from the interface and then disappear, after some time when the stresses are relaxed. It is most probably that in this transitional case the flow in the stretched bridge is influenced locally, also near the interface, by the nucleation of micro bubbles. Even if the size of the bubbles does not exceed the critical diameter of the cavity formation, they can still influence the flow near the moving meniscus.  

\begin{figure}
    \centering
 \setlength{\figW}{6. cm}%
    \setlength{\figH}{4.2 cm}%
\input{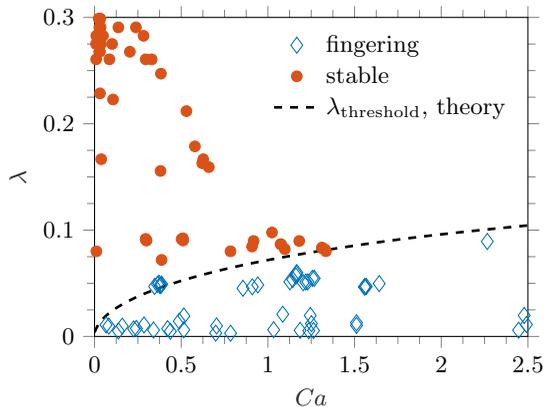}
\caption{Nomogram for the outcomes of the liquid bridge stretching for various values of $\lambda$ and capillary number $Ca$. The threshold for the bridge fingering is obtained from the full computations for $Re=0$ of $\lambda_\mathrm{threshold}$ corresponding to the condition $N_\mathrm{max}=5$. The approximate solution (\ref{eq:lam_thr}) is also shown on the graph, but it is indistinguishable from the results of full computations for $Ca<1$. } \label{fig:n_t}
\end{figure}

In fig.~\ref{fig:n_t} the outcomes of the liquid bridge stretching (stable receding of the meniscus without fingering or emergence of the apparent fingering) are shown for various values of $\lambda$ and $Ca$. It is not always easy to determine the outcome at the limiting cases near the threshold conditions. In this study the fingering is identified if more than five periods of the appeared interface waves can be clearly observed. The condition, $N_\mathrm{max}=5$, is used as a criterion  for the selection of the experiments leading to fingering. This criterion also allows theoretical prediction of the threshold value $\lambda_\mathrm{threshold}$ for  given capillary and Reynolds numbers. As shown in fig.~\ref{fig:Nmax_scaled}, the influence of the Reynolds number on the number of the fingers is minor and as a first approximation the threshold value $\lambda_\mathrm{threshold}$ is a function only of $Ca$. For small Capillary numbers the threshold value of $\lambda$ can be estimated using the approximate solution (\ref{eq:NmaxSmallCa})
\begin{equation}\label{eq:lam_thr}
   \lambda_\mathrm{threshold}\approx 0.076 \sqrt{Ca},\quad Ca \ll 1, \quad Re =0.
\end{equation}

It is interesting to note that in the limit $Re=0$ the same scaling as in (\ref{eq:lam_thr}), namely $\lambda_\mathrm{threshold}\sim \sqrt{Ca}$, corresponds also to a certain amplitude $G_\mathrm{threshold}\approx 1.7\delta_0$ of the shape perturbations $\delta(y,t)$, where $\delta_0$ is the initial shape disturbance. This relation can be obtained from (\ref{Gstar}). Since the initial disturbance $\delta$ is very small, the perturbations of the amplitude $G_\mathrm{threshold}$ cannot be resolved with our optical system. Note however that $G_\mathrm{threshold}$ characterizes the amplitude of the perturbations at the predicted time $\tau = \tau_\star$ corresponding to the  maximum number of the fingers. This amplitude continues to grow nearly exponentially in time. This is why in the cases close to the threshold, the fingers can be recognized at times slightly larger than  $\tau_\star$ and thus at radii close to $R/R_0 \approx 0.8$, as shown for example for the case $Ca = 0.29$ in Fig.~\ref{fig:functionRR0}.
 
Therefore, both conditions, a certain number of finger and a certain amplitude of the disturbances, can be used as  conditions for the observable generation of fingers. 

\section{Conclusion}

In this study the pattern formation in a liquid bridge stretched by an accelerating substrate is investigated experimentally and modelled theoretically. The maximum number of fingers is measured for a large range of liquid viscosities, gap widths and substrate accelerations.  

The process of finger formation is studied using linear stability analysis for small perturbations of the liquid bridge shape. The theory accounts for the viscous stresses, capillary forces and inertial effects. The model is developed for a single-sided accelerated substrate. It allows calculation of the amplitude of a certain wavelength on the bridge surface over time. Consequently, a prediction was derived for the number of fingers. The agreement with observations is good, despite the fact that no adjustable parameters have been introduced into the model. The prediction is however only applicable if no cavitation occurs. For experimental cases where cavitation occurs, the theory overestimates the number of fingers.

A criterion $\lambda_\mathrm{threshold}\approx 0.076 \sqrt{Ca}$ has been obtained for the onset of  the fingering instability. The fingering assumed a certain number of observable fingers. For lower numbers $N_\mathrm{max}<5$ the instability is perceived more as the loss of the asymmetrical shape, but not fingering. For $N_\mathrm{max}<1$ the flow should be stable. 

An alternative condition for the fingering, namely the threshold value for the amplitude of perturbations, leads to the same scaling for the threshold conditions, $\lambda_\mathrm{threshold}\sim \sqrt{Ca}$. 

\section*{Acknowledgements}
The authors kindly acknowledge the financial support by the German Research Foundation (DFG) within the Collaborative Research Centre 1194 ”Interaction of Transport and Wetting Processes”, Project A03.

\bibliography{finlit.bib}
\bibliographystyle{jfm}

\end{document}